**Understanding Physical Processes in Describing a State of Consciousness: A Review**

Charles Johnstone [a] and Prashant S. Alegaonkar [b*]

[a]Department of Natural Sciences, Mbeya University of Science and Technology, Mbeya,

Tanzania

[b] Department of Physics, School of Basic Science, Central University of Punjab, Bathinda

151001, PB, India

**ABSTRACT**

The way we view the reality of nature, including ourselves, depend on consciousness.It also defines the identity of the person, since we know people in terms of their experiences. In general, consciousness defines human existence in this universe. Furthermore, consciousness is associated with the most debated problems in physics such as the notion of observation, observer,in the measurement problem. However,its nature, occurrence mechanism in the brain and the definite universal locality of the consciousness are not clearly known. Due to this consciousness is considered asan essential unresolved scientific problem of the current era.Here, we review the physical processes which are associated in tackling these challenges. Firstly, we discuss the association of consciousness with transmission of signals in the brain, chain of events, quantum phenomena process and integrated information. We also highlight the roles of structure of matter,field, and the concept of universality towards understanding consciousness. Finally, we propose further studies for achieving better understanding of consciousness.

## 1. Introduction

Currently, number of disciplines like philosophy, religion, physiology, neurobiology medicine, physics mathematics and computational sciencesare pursuing studies to develop understanding about the human consciousness. The subject is complex, multidisciplinary, and open ended.A systematic investigation would lead to enrich the knowledge about the consciousness thereby developing new insight and perspective of the human (subject) existence. There aremultiple definitions of consciousness developed in context to different approaches taken. However, at a fundamental level,the consciousness of a subjectis a state of wakefulness to perceive and interact with the surroundings. This alsoinclude state of sleep and coma. Moreover, consciousness is defined asan experiencerepresented in the form of content that has a temporal sense. It is a kind of private entity of an individual subject referred to as qualiai.e.a subjective experience[1].Consciousness,also,has a correlation with the mind. Speaking in a broader sense, it isa type of psychological state and related processes that could be connected to the state of consciousness in a complex fashion. A state of spontaneity that an individual subject possess is also termed as a self-consciousness. Consciousness has associated with the awareness, understanding, and intelligence of a subject[2] which indicates thatthe consciousness must have manifested together with the very first life. Since the awareness of the surrounding environment is essential for the survivalof a subject to distinguish the sources of food, energy, hazardous things,and extinctions[3]. Another notable definition proposed by Grandy and Buffalo [4]states thatthe consciousness occurs due to theinteraction of a subject with the external environment in which one type of energy interacts with another form of the energy. This definition warrants the existence of more forms of the consciousness apart from the human consciousness such as atomic, chemical, and molecular consciousness. However, therea stillan open debate about existence of consciousnessin other creatures or physical systems apart from the subject.

The subjective reality of the naturedependsuponthe consciousness [5], conversely,consciousness is unitarily subjective,experiential and manifest in the form ofuniversal existence. Ultimate quest is: how consciousness originates in theuniverse? what is its mechanism of emergence? Hameroff and Penrose [5] identified three possible origins: (1) an evolutionary biological adaption of the brain and nervous system: in this case consciousness is considered as epiphenomenal,universally non-intrinsic, and has no distinct role [6]–[9]. (2) Dualism and spirituality: according to the spirituality consciousness has been in the universe all along and canonically influences all physical matter and human behavior [10, 11]. Dualism outlinestwo distinct realms of existence; the material universe that we perceive around us and the non-physical universe that is related to the conscious mind. However, in recent years this assumption is losing its popularity among many scientists since it violates the law of conservation of energy[12]. (3) Consciousness as the physical events: These events exist in the universe as '*non-cognitive event* or *proto-conscious event'*[5]. They can be transformed into consciousness byenhancing various conditions and physiological activating agents [13]. For example, in neuroscience and biology these events are associated with neuronal activitywhich gives rise to conscious moment.According to this approachconsciousness is consideredas a fundamental feature of the universe and it can be accounted by known or yet known physical laws of the universe.

There are number of enigmatic features of consciousness whose description isstill unclear.These includes; (1) the nature of the subjective experience; why are we always under a subjective experience?, (2) the binding problem; which refers to how distributed activities of the brain are unified into one content, (3) free will or non-computability, (4) non-locality[14].Further, as discussed earlier, Hameroff and Penrose [5] have highlighted that the nature, mechanism of occurrence in the brain, and definite universal locality of the consciousness are not clearly known. In this regard, consciousness is considered to be an essential unresolved scientific problem of the current era[15].According to Max Tegmark consciousness is like an elephant in a room[16].This argument implies that, consciousness is the biggest sensitive problem waiting for the solution. This claim brings an attention of the scientific community to explore consciousness as a valid area of the scientific research.

A better understanding of consciousness requires both empirical studies and a theoretical approach. The empirical studies can be used to gather informationabout the consciousness and corresponding theory could provide the behaviour of consciousness beyond the obtained empirical data. To give an example, why does consciousness disappear during aesthesia while the brain activity is intense and synchronous? Why does itdie out in early sleep where the brain is still active? And why it is relatedwiththe cerebral cortex and not with the cerebellum with its complex neuronal network[17]?

Further, the good understanding of human consciousness would lead to the fresh treatment of brain injuries, phobias, and a deeper understanding of mankind[18].One of the step towards understanding consciousness is to revisit the question raised by Schrodinger about; what physical process or event is associated with consciousness [19]? Our review focuses on addressing this question. Towards achieving this, we explored the association of consciousness with transmission of signals in the brain, chain of events, quantum phenomena process and integrated information. We also discussed the roles of structure of matter, field, and the concept of universality towards understanding consciousness. Throughout our discussion we have highlighted the open questions which requires further attention of researchers.

## 2. Association of consciousness with transmission of signal in the brain

The brain consist of about $10^{12}$ billion nneurons, 100 trillion synapses and $10^{13}$ billion glia cells, and each neuron can be connected to the thousands of others neurons[20]. Due to this,the brain is considered as one of the most complex systems in the universe. The human brain also appears as the system of network with high structural organization. It functions as the super-system that integrates different organizational levels, ranging from the neuron and the synapses, to a local cortical circuit and subcortical nuclei, to large-scale network. And the active communications between all these parts of the brain which can be mediated by the chemical or electrical mechanism of information transmission is also known as the brain activity.

The key question which we want to discuss here is, can the brain generate consciousness? To defend this question neuroscientists,posit that consciousness arises from the neuronal computational of the brain networks. The neuron can receive, process, and transmit information. It is well-accepted that neurons communicate with each other's, at the synapses using the chemical or electrical mechanisms. Chemical transmission is mediated using the neurotransmitters, where the electrical operates by passing current from presynaptic neuron to postsynaptic neuron. The process of opening of the channel in the presynaptic vesicular grid and discharging neurotransmitters into the synaptic cleft is known as exocytosis.

The electro-chemical signal transmission in the brain provides a satisfactory explanation of the low-level neural function such as sensory, reflex,and motor. But it fails to provide a reasonable explanation for higher brain functions such as consciousness, emotion, learning, and perception. This is due to the number of spikes fired by the neurons generated from the electro-chemical transmission which are considered as the mechanism of encoding neural information, their fire rate are not fully correlated with the neural function, because sometimes are very space or silentunder the appropriate behaviour condition[21], [22]. The other reason is about the low speed of action potential transfer in the neuron limited to about 120 m/s and it even more lower across the chemical synapse. It is difficult to account the rapid change of conscious state by this speed. This challenge has led to the proposal of other mechanism for information transmission in the brain. Among these include an attempt to use quantum tunneling, which is discussed in the subsequent section.

Towards resolving the inability of electro-chemical signal transmission to account for the consciousness, Beck and Eccles [23] proposed the quantum model for the release of neurotransmitter at the synapses in the cerebral cortex. According to this model, the quantum tunneling of quasiparticles which triggers exocytosis is related with the influences of consciousness.The tunneling process of quasiparticles was characterized by two energies [23]; thermal energy, $E_{th} = \frac{1}{2} k_B T$, where $T$ is temperature and $k_B$ is Boltzmann constant and he quantum mechanical zero-point energy $E_0 = \frac{(\Delta p)^2}{2M} = \left(\frac{2\pi\hbar}{\Delta q}\right)^2 \frac{1}{2M}$ , where $M$ is the mass of the particle, $\Delta p$ is momentum, $\Delta q$ is localized distance and $\hbar$ is planks constant divided by $2\pi$. The borderline was established as $E_0 = E_{th}$, $E_0 \gg E_{th}$ for quantal and $E_{th} \gg E_0$ for the thermal regime. The critical mass of quasiparticles $M_c$ were determined by taking fixed values for $T \approx 300K$ and $\Delta q \approx 1$A, these yields $M_c \approx 10^{-23} g = 6M_H$ where $M_H$ is the mass of hydrogen. This indicates that a quantum particles whose mass is less $6M_H$ undergo quantum tunneling across a potential barrier by triggering exocytosis [23]. This estimate also show that the trigger mechanism occursin atomic process and is not a temperature sensitive. However, this predictions was later disproved by experiments *in vitro* as vesicle fusion driven by neuronal protein machinery is arrested at 4 °C[24].          `

Georgiev and Glazebrook [25] have extended Beck and Eccle model to a molecular basis by identifying the quantum quasiparticle as Davydov soliton (DS). The DS acts as the twist for protein α-helices and triggers the synaptic transmissionacross the helical zipping of the SNARE complex, a protein molecule for vesicle fusion. According to this model, the mass of DS is about 5% of the hydrogen mass [25], [26], this is much less compared to the $6M_H$ obtained by Beck and Eccle. Since DS has a small mass, it can tunnel through a potential barrier of 1-2 *nm* thick[25]. It is similar to the conformational changes of the SNARE complex that are needed for synaptic fusion [27]. The other significant feature of this model is its ability to accommodate temperature dependence [26]. Where the thermal oscillations of the potential barrier increase the probability of tunneling. The model also supports the physics of consciousnessat the biomolecular level, especially in the neurotransmission process via the SNARE protein complex which induces exocytosis in the synaptic cleft. Finally, according to this model, the quantum chemistry of the SNARE complex can be considered as a substance for further investigation in association with the natureof consciousness.

Another promising alternative mechanism for neural transmission and processing of signalthe brain which may in future will provide an account for consciousness, is  an attempt to use biophoton[28].These are ultra-weak photon emission (UPE) from living organism, emitted near UVA,visible, and near IR in the spectral range of 350 to 1300 *nm*and their intensity ranges from $10^2$-$10^3$photon/(cm$^2$s)  [29]. Some studies have shown that biophoton activities can be transmitted along the axonal fiber of neural circuit [28], [30].However, in order to achievea better understanding of biophoton transmission in the brain, we need to understand; the origin of the biophoton signal in the brain, the role of biophoton in neural communication and informational processing, the mechanism by which bio-photon is transmitted across the neural circuit, and how to construct a model for bio-photonic transmission[31].

### 3. Consciousness as the chain of events

The argument of consciousness as the discrete of events has a long history. It can be traced back to the time of William James [32] who formulated 'specious present' doctrine of temporal experience. It shows the content of one's perceptual experience spans a time-based interval.According to Stroud[33] the discrete events of consciousness occur like movies frame at 24 to 72 frame per second (24 to 74 Hz). Buddhist text quantify conscious moments as 6,480.00 in 24 hours (75 Hz) for Sarvaastivadins while for some ChineseBuddhists is one thought per 20 ms (50 Hz) [34].For neuroscientists the best correlate of consciousness occurs at gamma synchrony (30 to 90 Hz) electroencephalography (EEG)[35], [36]. This implies that consciousness consists of discrete events occurring within brain regions at varying frequencies. It can be 40 to 90 conscious moment per second.The source of these events is associated with cortical and thalamocortical oscillation in different frequencies bands. They can be used to provide the constrained framework for investigation of computation leading to awareness [35]. Moreover, the subjective nature of our experiences suggests that neuronal events evolve continuously. This creates an open question whether we experience the world as continuous signal or discrete sequence event.

Schrodinger proposed that, the sensations, perceptions, and memories are the construct from which the universe is composed of. The consciousness experienced by an individual subject relates to the unitary consciousness manifested universally. It has been proposed by Schrodinger that, under specific conditions consciousness manifest in the brain. However, what enables such manifestation? does it depend on the physical property of brain? what kind of material process is associated with consciousness? is still unclear. Schrodinger was of the opinion that, from the experience of an individual subject, consciousness is connected to the specific chain of events in the organized active matter of the brain through nervous function. It involves the mechanism by which the subject responds to periodic and aperiodic events for adaption of the changing surrounding.

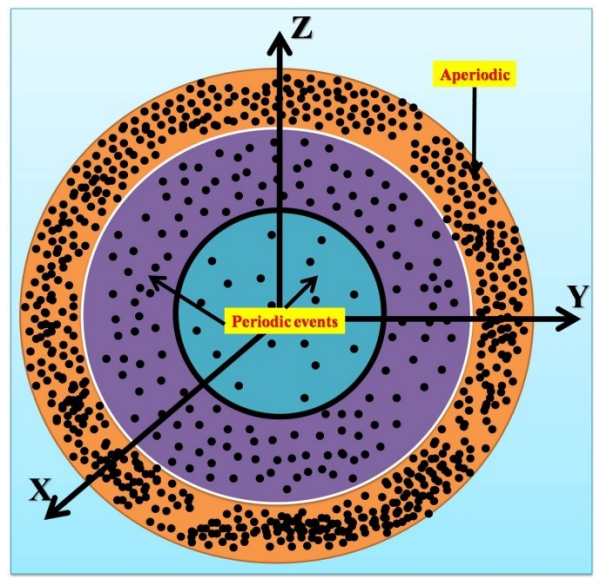

Figure I: Schematic representation of sphere of consciousness envisaged by Schrodinger showing density of periodic and aperiodic events contributing to response of a subject.

Further, Schrodinger departs from the western philosophical model of consciousness put forwarded by Spinoza, Gustav Theoder Fchne [19]. Schrodinger illuminated the idea of sphere of consciousness that distinguish events in which aperiodic event are on the surface of sphere as schematically shown in Figure I. The periodic events are fade and away from surface of consciousness sphere. Schrodinger also speculated the relevance of quantum phenomenon to the bio-systems which can be extended to consciousness as discussed in subsequent section.

## 4. Consciousness as the quantum phenomena process

Quantum mechanics,as afundamental theory in Physics remarkably explained the behavior of matter at an atomic and sub-atomic level.In recent years, number of biological processes such as photosynthesis, bird navigation, and olfactory reception have been iterated using quantum physics[16–20].This has paved a notional foundation for the application of Quantum Mechanics in describing Bio-systems.Probably the agreement betweenthese two fields can be attributed to the size and the scale of the objects.

Earlier, the use ofQuantum Mechanics to explain the behaviour of life has been discussed by Erwin Schrodinger [19]. He proposed that, human life has a tendency of heredity which has a molecular basis.This was later confirmed as the DNA: the genetic basis of the human evolution [42].Thus,interfacing Quantum Mechanicswith Bio-systems has two-foldfocus: (1) can bimolecular behavior beinterpreted quantum mechanically? (2) are nontrivial quantum phenomena relevant to life? [37].

The greatest challenge thatremained unaddressed is to explain how consciousness emerges from the structure, processes, and function of the brain which is a bio-system thatoperates at a physiological temperature under certain physical laws. At a fundamental level, it can be argued that, all bio-systems are the quantum systems, composed of molecules and can be implemented forthe quantum mechanical effects, especially, superposition, coherence, entanglement, tunnelingfor accountingconsciousness[23-24]. This view has been surveyed subsequently.

*(a) Collapse of the wave function*

The wave function represents the probability of finding a particle at a location in a quantum system. Moreover, it has a property of superposition that shows diffraction and interference i.e. to co-exist at multiple places. Such a superposed quantum state reducesto a single state or collapse to a classical statewhen the measurement has been performed, which is termed as the collapse of a wave function.It is referred as the measurement problem. The wave function obeysSchrodinger's formulation which is indeed linear and predictiveaboutits behavior, however, itdoes not account for the measurement. The nonlinearity encounters due to the state of measurement which isdiscontinuous, and non-deterministic. Several attempts have been made to overcome the measurement problem. The conventional view is that of Copenhagen interpretation where the quantum states are reduced by measurement, environmental entanglement, and conscious observation (i. e. subjective reduction SR or R). However, in this article, we will restrict to those approaches which are associated with the consciousness.

One of the approaches is Von Neumann and Wigner who hypothesized that consciousness causes a collapse of the wave function.Von Neumann [45] proposed that, the mathematics of quantum mechanics permits the collapse of the wave function to be located at any position in the causal chain from the measuring device to the human perception of the measurement.He carried out the treatment of relating Quantum Mechanics to various causal and statistical method of describing nature of consciousness. Neumann pointed out a peculiar dual nature of Quantum Mechanics procedure that could not be satisfactorily explained.He pointedout that, the transformation of a state φ to $\varphi'$ under the action of energy operator generate a pure causal state given by φ'= φ′ = e$^{(- i/ℏ\ t\ H)}$ φ. However, such a state is a mixture of unitary state given by:$U' = e^{(-\frac{i}{ℏ}tH)} U e^{(+\frac{i}{ℏ}tH)}$ as anoutcomeof causal change.The state can be represented with respective probabilities which is not the process resulted from the causality. There is the fundamental difference between $U$ and $U'$ in terms of reversibility in which later is irreversible.

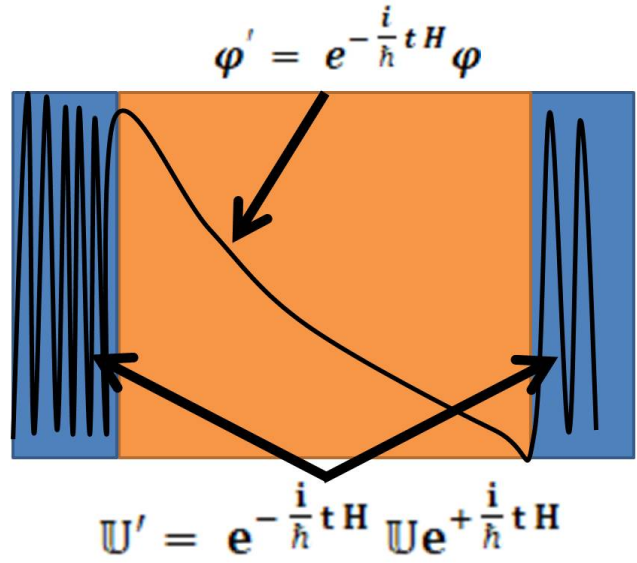

$$\varphi' = e^{-\frac{i}{\hbar}tH}\varphi$$

$$\mathbb{U}' = e^{-\frac{i}{\hbar}tH}\mathbb{U}e^{+\frac{i}{\hbar}tH}$$

Figure II: Schematic representation of psyco-physical parallelism proposed by John Von Neumann in which evolution of the quantum state φ and φ′ is shown indicating mixing of the states of U and $U'$.

He pointed out a basic requirement of the scientific viewpoint in the form ofa psycho-physical parallelism. According to him, such mechanism enables toprovide additional physical inputs of the subjective perception attributing the apparent feature of reality of the physical world. In other words, this is equivalent toallocate physical processes in the objective environment, in ordinary spaceled to conscious experience, as shown schematically in Figure II.

Based on the Von Neumann's chain, Wigner proposed that the measurement is completed when consciousness is attained [46]. According to Wigner's friendexperiment, measurement isassumed to be the interaction that creates impressions in our consciousness,thereby causing modification of the wave function of the system. This assumption is in agreement with that of London and Bauer [47]argumentwhich states that consciousness completes the process of quantum measurement. According to this claim, no physical process causes the collapse of the wave function. The task of the state reduction is directly linked to the consciousness. However. it is unclear that,how consciousness can cause the state reduction of a physical system? In later years, Wigner by himself dismissed this proposal due to the two reason: (a) physical reason; macroscopic objects can never be considered as an isolated system, which means quantum mechanics does not applytothe description of the behavior of macroscopic bodies, (b) philosophical reason; the implication of solipsism on physical reality and the interpretation of the wave mechanics [46]. His idea was maintained by Stapp [48] who argued that subjective reduction of the wave function collapse in neurons is associated with the consciousness exist in the brain. This approach assumes that measurement is an act of the consciousness. This approach implies an attempt to solve the consciousness-causation and measurement problem simultaneously. Where,correlation betweenconsciousness and measurementis still unclear. However, there are certain motivations behind this approach which include; measurement by a conscious observer, consciousness-causation problem, and conscious observation have definite results.Despite these motivations, the approach is unpopular, this might be due to theobjection raised against it which are impression and dualism[49].Surprisingly inrecent years, there has been a proposal to combine quantum collapse dynamics withthe mathematical theory of consciousness such as integrated informationtheory (IIT)to yield a precise interpretation of the consciousness[49]. The focused question of this view iswhether can a collapse model be defined by the quantitative measure of consciousness, such as the Phi ($\Phi$) measure of IIT.The assumption behind this is that $\Phi$ resists superposition and the superposition of $\Phi$ triggers collapse. The model developed on this basis is affected by the quantum Zeno effect and lack of precise symmetries between brain states associated with the different experiences [49], [50].Yet this model is open for the empirical testing, which implies that the consciousness-collapse model needs further attention of the researchers[49]. Particularly,invariable-locus models where special properties serve as the locus for the collapse, and alternative models where the physical correlate of the consciousness encompasses complex wave function property, or which involves the independent variation of consciousness with any physical properties.

Another approach is that of the multiple-world interpretationwhose foundation lies in Hugh Everett's doctoral thesis [51]. According to this view, consciousness does not cause the collapse of the wave function, instead, it results in the branching of the universe into a parallel universe.For example,the case of the Schrödinger cat experiment results in two parallel universes, one in which the cat is alive and the other in which the cat is dead. These branches of the universe are associated with the outcome of measurement and are considered as the subjective experience of the corresponding observers. But what causes the split of the universe? It is still unclear, due to the uncertainty in measurement process [52]. Further, the formation of multiple universes violates the conservation of mass-energy [53].

*(b) Orchestrated objective reduction theory*

Another remarkable application of quantum mechanics in consciousness is that of Penrose and Hameroff in their orchestrated objective reduction theory (Orch OR). In this theory, consciousness is considered as the result of coherent quantum superposition of tubulin of microtubules which terminate based on objective reduction (OR)[2]. Each superposed state has its space-time geometry, as the degree of coherence mass-energy differences which attain the threshold of separation related to quantum gravity, at this point the system reduces to the single state, termed as the ''conscious now'' [54]. It is attained at $\tau \approx \frac{\hbar}{E_G}$, Where $E_G$ is gravitation self-energy of superposition, $\hbar$ is Planck's Dirac constant and $\tau$ is coherence time. This kind of reduction of quantum state to classical is interpreted as the conscious perception and it generates a particular pattern of the microtubule-tubulin conformational state that regulates neural activities[14]. Due to this, Orch OR theory operate at the interference between classical neurophysiological and quantum mechanical forces. Furthermore, the time taken from superposition to collapse is associated with pre-conscious while the time at the collapse is related with the point at which transition from pre-conscious to conscious occur. The conscious perception which occurs in this situation is non-computability (non-deterministic), this is due to the self-collapse of OR which is related to quantum gravity[5].

In Orch OR theory, quantum entanglement act as a means of interconnection of tubulin qubits that are superposed with other superpositioned tubulins in the microtubules lattice. This enables the superposition of microtubule tubulin of many neurons, which allows unity and binding of conscious content.Furthermore, Consciousness interconnect in the universe between the quantum holographic brain to the quantum holographic space-time reality by means of a quantum information entanglement [55]. Here, space-time is considered as the manifestation of quantum information entanglement [56], [57]. Quantum entanglement is taken as the form of information which prevails in the space-time. By this argument consciousness can also be associated with structure of the space-time geometry [5]. This approach allows the extension of consciousness into space-time contrary to Descartes arguments that mental entities are un-extended[58]. But how consciousness depends on the existing space-time or does it have its own space-time is still a matter of investigation.

*(c) Limitation of quantum physics towards addressing consciousness*

The Orch OR has received several critics since its inceptions. According to Tuszynski [15] the critics can be grouped into three categories; (1) Lack of experimental evidence linking between single synapse and dynamics of neural assemblies activity. Thus, experiment validation is needed to verify the relevance of the quantum process to cognitive processes. (2) currentlythere seems no quantum mechanical properties to describe psychological and neurophysiological phenomena. (3) an irrelevance of biological structure to quantum effects, it argued that large structures such as neurons and microtubules do not support quantum effects, since they function at high temperatures. The high temperature in the quantum system causes decoherence and eliminates the possibility of quantum effects playing roles in the brain process. Our discussion focuses on the critics based on feasibility of quantum effects in microtubules. Since it is one of the maincomponents of the theory apart from the quantum gravity. The aspect of quantum gravity is open for testing[59],but the problem is the technological barrier [60], [61]. However, a recent attempt of quantum formulation of Einstein equivalence principle (EEP) has shown that it's validity of classical EEP does not apply for quantum formulation.[62].This suggests that the validity of EEP in quantum theory requires an independent experiment verification not inferred from classical experiment. The promising future experimental investigation should be centered on the condition where general relativity affect internal dynamics of low energy quantum system [62].

Microtubules have been suggested as the site which can support quantum effects. It serves different biological functions such as; cell division, cell movement, transport for macromolecular, and maintenance of cell form and function. However, the attractive features of microtubules for quantum effects includes; (1) the presence of aromatic amino acid. (2) short distance between aromatic amino acids in adjacent tubulin dimer ($\sim nm$) (3) the long-range order of the microtubule lattice (4) the directionally polarized and helical arrangement of tubulin in a microtubule [63]. Despite all these properties, the feasibility of quantum effects in microtubules has been criticized by Tegmark [64] who showed that the decoherence time scale ($\sim 10^{-13} - 10^{-20} s$) for microtubules are very shorter as compared to the relevant timescale for the cognitive process which lies on the order of ($\sim 10^{-3} - 10^{-1} s$). This drawback has been refuted by Hagan et al., [65] who argued that Tegmark didn't use the Orch OR theory conditions in his calculation and thus his proposal was irrelevant. By using the Tegmark formulae; $\tau \sim \frac{4\pi\epsilon_0 a^3 \sqrt{mkT}}{N q_e^2 s}$, Where $m$ is the mass of ionic species, $T$ is the temperature, $a$ is the distance from the ion to the position of the superposed state, $s$ is the maximal separation between the position of the tubulin mass in the alternative geometries of the quantum superposition and $N$ is the number of elementary charges comprising that state, Hagan and his colleagues recalculated decoherence time for microtubules by incorporating Orch OR theory conditions (superposition separation, charge versus dipole, and dielectric constant) and obtained ($10^{-5} - 10^{-4} s$) which is sufficiently close to the relevant neurophysiological range scale [65]. The highertemperature results into strong decoherence. As biological systems are wet and warm, for this reason they are liable environment for decoherence. However, it has been shown that biological system avoids decoherence by using several mechanisms, these includes; (i) isolation of the quantum system from environmental interaction by shielding/screening. For example, the microtubule is shielded by the counterion

Debye plasma layer and water-ordering actin gelation [66]. (ii) Thermodynamic gradient: it allows the biological system to act as a heat engine to reduce the effective temperature of certain molecular complexes. For example, the slow release of energy from ATP (Adenosine Triphosphate) molecules at the actomyosin complex, implies a quantum coherence on the macroscopic scale [67]. (iii) Decoherence-free subspaces: in the case of quantum computer building this is related to identifying sub-spaces of Hilbert space that are free from the coupling of the system with its environment. On the other hand, it occurs when the system freezes other degrees of freedom by sort of quantum Zeno effect due to being strongly coupled to its environment at a certain degree of freedom. This allows entanglement and quantum superposition to persists [68]. (iv) The structure of microtubules also contributes to the avoidance of decoherence since it is suited for topological quantum error correction [66]. Furthermore, the ongoing experimental testing of Orch OR theory of consciousness to detect presence of quantum process in microtubules and their sensitivity to an anesthetic under the project ''Effects of an aesthetic molecules on quantum vibration in microtubules'' is expected to generate new understanding of the theory at the nanoscale [69]. As recently, Stuart Hameroff [70] claimed that Orch OR has broad explanatory power and can be easily falsified by demonstrating the absence of quantum interference in microtubules, or if exist, proves it insensitive to anesthesia. This is one of the strongest supports for validity of the theory if will be proved experimentally.

The decoherence effects seem to be a major challenge for biological quantum systems and the quantum theory of consciousness. Although decoherence destroys quantum coherence, it may at the same time enhance the transfer of energy [71]. Studies show that the coherence lifetime for photosynthesis and radical pair mechanism is in Picoseconds and microseconds respectively [72]–[74]. These time scales are smaller compared to the regular neuron firing time scale which is in milliseconds. It appears the biological system extends beyond these time scale to attain the neuronal firing time scale for the cognition state, otherwise firing of nerves do not occur in this time scale. However, an open question remain as how the biological system attain this condition? The study by Fisher [75] using entangled Posner molecules propose the existence of longer coherence time for even a day. Rooting his study on the concept of Hu and Wu [76] who postulated that consciousness is mediated by spin quantum. He explored the feasibility of quantum cognition based on nuclear spin. Nuclear spin are weakly coupled to environmental degree of freedom, for this reason their phase coherence time lasts for five minutes or longer [77], [78]. Magnetic and electric field perturbation causes the decoherence of the nuclear spin. For this reason, nuclear with spin $I > 1$, the presence of quadrupole moment which couples with the electric field and in addition to the magnetic field produced by nearby nuclei cause quicker decoherence. Where for nuclei with spin $I = {}^{1}/_{2}$ are more weakly decohered only by magnetic field. Thus, the element with half nuclear are essential for hosting putative neural qubit [75]. However, the application of nuclear spin in brain is still lacking of realization of conditions such as.; biological qubit with long nuclear-spin coherence time, transportation mechanism of qubit, quantum memory storage of qubit at molecular scale, quantum entanglement mechanism in multiple qubits, and the chemical reaction which induces quantum measurement on the qubits [75]. The validation of these conditions will aid in supporting or denying the existence of nuclear spin quantum process in the brain.

*(d) Quantum brain dynamics*

The idea of Quantum brain dynamics (QBD) proposes that water which forms around 80% of the brain, rather than being passive, it could be an active player in the brain process. The electrical dipole of water molecules in the brain constitutes a cortical field. The cortical field contain energy quanta behaving as particles, which are known as corticons[79].Corticons are said to exist everywhere in the cerebral cortex and they interact with the main dynamic of neural network. This phenomenon gives rise to the transmission of signals within the body.

According to Ricciardi and Umezawa [80], memory is equivalent to macroscopic quantum state with long-range correlation. Based on this model, the brain is considered as the biological system equivalent for dynamic symmetry breaking.Its formulation is based on quantum field theory (QFT)[81]. The motivation for using QFT isbecause information in any materials is carried by ordered pattern, maintained by certain long-range correlation,and mediated by massless quanta[81]. The generation oflong-range correlation occurs due to the spontaneous breakdown of symmetry. It is responsible for generating and maintaining an ordered pattern (coherence) in the system. The vacua or coherent state are responsible for memory storage. The recording process of memory in the vacuum state is achieved by coherent condensation of Nambu Goldstone modes. These are massless boson particles that appear to be the dynamical response to the breakdown of symmetry, responsible for quantum mediating long-range correlation among the atoms. The symmetry breakdown occurs in QFT since there exist infinitely many ground states which are unitary inequivalent, contrary to quantum mechanics where all ground states are physically equivalent and thus do not support symmetry breakdown.

Jibu and Yasue [79] extended the idea of Ricciardi and Umezawa into a quantum theory of brain dynamics (QBD) to address the problem of consciousness. They hypothesized that the creation and annihilation generate consciousness. However, the authors do not give a reason as to why consciousness arises from this physical interaction and not from the interaction between electrical potential and chemical in the synapses.  According to their view, consciousness may be considered as a fundamental property of field, photon, or corticons. And it emerges from the interaction between the cortical field and other waves propagating along the neuronal network. The other extension of Ricciardi-Umezawa model as discussed in subsequent section are based on memory. As this approach continuous to contributes a better understanding of the mystery of memory, there is an open question as whether it canalso assist towards understanding consciousness.

The quantum brain model associated specific memory to a specific vacuum code. Once this is selected for printing certain information, there is no other vacuum state successively accessible for recording another information,unless external stimulus carrying new information creates new vacuum state by phase transition. This causes the destruction of the previously stored information (overprinting). To overcome this problem Vitiello [82] extended the model to dissipative dynamics, where infinitely many vacua are independently accessible. This allow hugenumber of information to be recordedwithout destroying the previous information, thus allowing a huge memory capacity. To achieve this an assumption is made to break both the rotational symmetry and time-reversal symmetry before information recording process[82]. This makes the brain state to be completely determined after information has been recorded. And it cannot be takenback to the previous state.This fact is referred to well-known warning ''...*NOW you know it!...*'', which means once you know, you becomedifferent from the previous time. This introduces the *arrow of time* into brain dynamics; the distinction between the past and the future based on the information recording.

Another development of the quantum brain model is that of Pessa and Vitiello [83] who extended the model to the role of entanglement, quantum noise, and chaos. In their model, they have shown that doubling the degree of freedom of the system accounts for quantum noise in the fluctuating random force in the system-environment coupling. Since brain-environment entanglement is permanent, this implies that quantum noisy effects are intrinsically and inextricably present in the brain dynamics. They have also shown that trajectories in memory space may exhibit chaotic behavior. This may account the high perceptive resolution in the recognition of the inputs.The recent extension of a quantum field theory of the brain is that of Nishiyama et al [84]whom they have extended the model to non-equilibrium electrodynamics in open system. The main assumption of this extension is to introduce the non-equilibrium multi-energy-mode analysis in open system which lacks from the presented models above. This would assist to demonstrate whetherdecoherence, the main criticism of QFToccurs or not in the open system. Towards achieving this, they derived Klein-Gordon equations and Kadanoff-Bayem equations to describe the non-equilibrium, charge-energy conserving, and entropy-producing dynamics. These equations can be applied to analyze microtubules coupled to water battery surrounded by biochemical supply and for information transfer between two coherent regions via microtubules[84].

## 5. Consciousness as an integrated information

The human life always involves transmission of information such as genetic information, verbal information, historical information, digital information, or written information. It seems life and information are inseparably united. Information always carries a physical substance like nerve impulse in the brain, sound wave in air and electromagnetic signal in space. Further, according to Chalmers[85] information has two aspects, physical one and awareness. Awareness is taken as the property of information like mass as the property of matter or frequency as the property of field. As we discussed in introduction section consciousness is associated with awareness, it can also be equated with information. Hence the substrate of consciousness must possess the ability of encoding information.Information is the fundamental property of the universe, it is found everywhere, not limited to human brain, living matter or space-time[86]. Due to this, equating consciousness with information imposes the difficult to account for the phenomenon at all.If consciousness is everywhere, why objects apart from the human brain do not exhibit conscious? To resolve this, a further approach is proposed to equate consciousness with information that exist only in certain dynamics states [87]. Based on this assumption not all information's are conscious. This introduces another question; what are the dynamic features which distinguish a conscious information from unconscious one? In the next subsequent section, we discuss the application of this approach based on integrated information theory.

Integrated information theory (IIT) is the mathematical theory of consciousnessthatrelate consciousness withintegrated information (Phi Φ) [88], [89]. In this approach, consciousness is considered as the fundamental quantity like mass, charge, or energy. This allowsfor study of the subject on what physical laws or properties it obeys, just as physicists have studied new forces, fields, and particles in the past. Based on this regard, IIT proposes five axioms of consciousness; (1)existence; consciousness is a real and undeniable fact. However, its existence is intrinsic, it exists from its own perspective. (2) Composition; it is structured such as red, a tree, a table, a book, etc. These structures allow for distinctions of physical objects of the universe. (3) information;it is specific, each experience differs from many other possible experiences. (4) integration;it is unified,irreducible to the non-interdependent component.For example, experiencing a blue book is irreducible to seeing a book with no blue color, plus a blue color patch but no book.(5) exclusion; it is definite in terms of content and spatiotemporal grain. It has borders, excludes others, and also flows at a certain speed[88], [89]. Are these properties sufficiently enough to account the phenomena of consciousness? The answer actually is No,they are not sufficient [90].IIT accept this by upholding that all its axioms are self-evident [89], [91]. They can be used to further understanding of consciousness. IIT transforms these axioms into the language of physics to check whether the physical substrate of consciousness satisfies these properties. It uses the construct of mechanism (logic gates, neurons-like) and system (computer, neural architectures) to validate the postulates of the axioms.With an assumption of the physical substrate of consciousness must possess maximum intrinsic cause-effect power[17].

IIT offer severalpredictions. Some of these include; (i)the loss and recovery of consciousness are associated with the breakdown and restoration of integrated information in the brain respectively[92]–[95].The loss of integrated information in the brain corresponds to unconsciousness states such as dreamless sleep, anesthesia, and comma. For example, consciousness vanishes in an anesthesia, because anesthetic molecules cause function disconnection in posterior complex thereby interrupting the cortical communication and hence generates loss of integrated information.(ii) Brain lesions; it causes unconscious if it disrupts the capacity for information integration. Clinician face a challenge to measure the consciousness of injured-brain, unresponsive patient based on the subject to interact with environment.To resolve this problem IIT suggests the measure of level of consciousness to be assessed based on the Φ dominant conceptual structure.The perturbation complexity index (PCI)measure of level of consciousness using transcranial magnetic stimulation (TMS) data set in wakefulness, sleep, and anesthesia, decreases in all condition of loss of consciousness[96]. This approach shows promising results for the application of IIT in analyzing empirical data of consciousness. It includes characterizing various states of consciousness in human brain such as sleep, anesthesia, and coma. This is promising technology for doctors in the future for detection of whether patient are conscious or not, especially for patient suffering from 'locked-in' syndrome, a condition where a patient cannot move or communicate[16].(iii) Consciousness is associated with only certain regions of the brain. For example, the cerebellum with an enriched network of neuronal connections does not give rise to consciousness while the cerebral cortex is associated with the generation of consciousness. This is because the cerebral cortex comprises elements that are functionally specialized and at each time interact rapidly and effectively[88], [97], [98]. This type of organization is associated with a high $\Phi$ value.(iv)The length of experience is correlated with the time interval at which its relevant physical element attains a conceptual structure with the highest $\Phi$ value. In other words, it is associated with the discrete-time interval at which cause-effect power attains a maximum.

Despite all these predictions, IIT suffers from different limitations. Itsformulation is based on the discrete with the finite number of states, where a physical system such as the brain change continuously, for example, the position of particles can take any of an infinite number of values [99], [100]. Applying the IIT formula to such a system yields $\Phi$ infinite which is unfeasible result. It also requires explosivecomputational for estimating the $\Phi$ value for a real system, this makes its application to be impractical for the brain[89], [101].The other criticismis the lack of empirical evidence for its formulation, non-functionalism, unclear quantification of consciousness, and definition of consciousness [90], [91], [102]. However, recently there has been an attempt to resolve the computational cost of IIT by the use of searching algorithms for minimum information partition and measuring $\Phi$ from high-density electroencephalography [101], [103].The successful of this attempt is expected to increase the applicability of the theory.

6.  **Consciousness as the structure of matter**

Despite of stronglyassociation of consciousness with the physical matter of the brain, there still unclear how consciousness emerge from the operative structure of matter and their processes[104]. The realization ofspecific operative structure of matter associated with consciousness would provide a ground for its existence and prediction. As the study of physics, biology and neuroscience deals with material structure and their processes at different level, can assist us to tackle the problem.It is expected these subjects could provide an integrated ground structure forthe validity, interpretive and predictive power of consciousness.The relation of consciousness with matter can be viewed into two aspects; (i) human consciousness is a continuation, so its origin is similar to that of matter and life, and (ii) complex dynamic physical organization of operation of matter and brain are related with consciousness [104].Based on these two aspects we can search for the possible physical substrate of consciousness from inorganic matter, organic matter and life, molecular metabolic, global regulatory system, and nervous system.

According to Tegmark [105]consciousness can be understood as the state of matter, with distinctive information processing abilities. This approach treats consciousness as an emergent phenomenon like solids, liquids, and gases.In physics, these emergent properties are measured by quantities like viscosity, compressibility. For example, when a substance is viscous, it termed as fluid, otherwise is solid.And if it compressible it a gas, otherwise it is a fluid.Can't we have properties like these to measure consciousness?Taking an example of emergent phenomena such as wetness, a drop of water is wet, while ice crystals and steam of cloud are not wet, though they are made from identical molecules of water.This phenomenon occurs since wetness depends only on the arrangement of molecules. In otherwords, wetness as emergent phenomena like solids, liquids, and gases have properties above and beyond the properties of their particles. It is also known as substrate-independent property. Likewise, what is crucial in consciousness alsois the particles rearranged.For example, being in different states of consciousness such as awake, dreamless sleep, and coma corresponds to different pattern rearrangement of particles[16]. These rearranged bunch of particles can be represented as the mathematical pattern in spacetime. Because mathematics equations of physics can describe patterns and regularities of the working of nature.As it has been proclaimed by Galileo Galilei nature is ''a book written in the language of mathematics'' and Eugène Wigner also has highlighted the usefulness of mathematics in natural sciences[106], [107]. Based on these assumptions, mathematics can be used to describe nature including consciousness. However, the implementation of this approach requires the identification physical parameters embedded within consciousness and its appropriate mathematical construct.

Tegmark [105] propose an alternative approach to the hard problem of consciousness as coined by Chalmers[85], to start with the *hard fact*, that is some of the arrangement of the particle are conscious while others are not. This approach raises one fundamental question; what properties of particle arrangement make the difference? This question can be resolved by establishing a scientific approach. For example,one can establish a theory that can predict whysome particles arrangementsare conscious and other are unconscious. This approachaffirms that consciousness is a scientific field contrary to what philosopher Karl Popper popularized that it is not a scientific field. Furthermore, taking Tegmark's[16] ideathat consciousness is the way information feels when processed in a physical system. And as discussed earlierconsciousness is the substrate-independent, due to this, the important property is the pattern of information processingand not the structure of the physical system.Therefore, consciousness isconsidered as an informational processing, then from this assumptionone can ask what properties are neededfor an information processing tobe conscious? By exploring this question Tegmark [16], [108]proposed four basic principles which distinguish a conscious matter from another physical system, these are (i) information principle; it mush has substantial information storage, (ii) dynamic principle; it refers to informational processing capability, (iii) Independence; it must differ from the rest of the world, and (iv) integration principle; it does not consist of nearly independent parts.His speculative on these factors gives insightful findings. For example,classical physics allows the integration of information about half of its bits by using error-correcting codes. Further, the information stored in the Hopefield neural network is naturally error-corrected, but the $10^{11}$ neurons support only 37 bits of integrated information.But the content of information of our experience are larger than 37 bits. This is known as an integration paradox[108].The generalization of these results to quantum information yielded more lower bits. Due to these, integrated principle needs addition information. Furthermore, investigation is needed to address whether; (i) does error-correcting codes existin the brain? (ii) can we find non-Hopefield neural network which can support more than 37 bits of integrated information[105]?Another finding is that of independence principle, where the best decomposition of Hamiltonian $\mathbf{H}$ by Hilbert space factorization is found in the energy eigenbasis, where $\mathbf{H}$ is diagonal. This resultleads to Quantum Zeno Paradox;it implies that when the universe is decomposed into maximally independence objects, then all changes grinds to halt.It produces a static world. As conscious observer does not perceive reality as static, due to this, independence principle also needs supplement additional principle.The interesting findingsis that of dynamic principle, where the energy coherence $\delta H \equiv \sqrt{2tr\dot{\rho}^2}$ was found to be a measure of dynamics. where $\rho$ represents the total density matrix.The best result was obtained by modest percentage reducing the $\delta H$ which enabled complex and chaotic dynamics, where increasing $\delta H$ was unable to support complex information processing. Finally, these principles are open for validation and can be used as tentative criteria for distinguishing conscious system from unconscious.

## 7. Consciousness as a field

In this approach, consciousness is considered to possess similar feature likethat of physical fieldsuch as ability to have duration and extension in the space. The origin of this idea can be traced back to the time of Gestalt-psychology in early twentieth century, where Köhler proposed that electric fields are cortical correlate of percepts[109].This assumption was contrary to atomist movement who argued that perception experience is the sum of sensory input. It emphasized perception is more closely related to field, rather than particle. The theory was disproved by Lashley[110]who showed short-circuit ofcurrent of visual cortex of monkey brain's do not cause any disturbance of visual functioning. However, in later years, the method adopted by Lashley's to disprove the theory was found to be inappropriate, it doesn'ttest whether disrupt affected vision-related current in the brain[111].Being influenced by Lashley's critics work,Libet[112] proposed that consciousness is a field, which is not in the form of known physical fields, he called it conscious mental field. It is difficult to observe this kind of fieldby directlyknown physical methods. It confines the theory into the realm of philosophy. Though Libet's theory proposes hypothesis for scientific testing, its success will not go beyond the electromagnetic field theory of consciousness discussed below.This theory struggled for several years to get publication dueto Lashley's legacy results[110].

The conscious electromagnetic field proposes that consciousness is the manifestation of the brain electromagnetic field [113].According to this theory, the massive neurons membrane depolarization generates electromagnetic field perturbations that influences the probability of firing of adjacent neurons.It has been also confirmed experimentally that endogenous field influences the brain functioning under physiological conditions[114]. Since neurons are densely populated in the brain about $10^4$ neurons/mm$^2$, so the adjacent neurons forms a complex overlapping of the field with the superposition of fields of millions neurons[115]. This feature represents the ability of the theory to integrate vast quantities of information in the single physical system. It also accountsfor the binding of consciousness. Further, the theory also proposes;

*"Digital information within neuron is pooled and integrated to form an electromagnetic information field. Consciousness is that component of brain's electromagnetic information fieldthat is downloaded to motor neuron and is capable of communicating its state to the outside world* [113]*"*

According to this theory, the superposition field is free from the influence of external field at all, due to the high conductivity of the cerebral fluid which creates an effective 'faraday cage' that insulates the brain from external fields [116]. But one can ask why the faraday cage do not deny the exit of generated electric field, which would suppress the possibility of recording electroencephalography (EEG) from the scalp? Further, why cannotradio waves with high frequency, main voltage with the same frequency as the oscillation proposed being conscious, and powerful magnetic field inside magnetic resonance imaging (MRI) affect the faraday cage? McFadden's [116] responded to these questions as follows; the source of EEG signals is the assemblies of neurons firing synchronously, not single firing neuron. This cause neurons to distribute and amplifies field effects. (a) The high frequencies of radio waves make them unfeasible to interact withthe low frequency brain-waves, (b) the electromagnetic oscillation at main voltage radiates ineffectively, hence no power is detectable, (c) the magnetic field inside MRI machines do not have right spatial configuration to couple with the putative conscious field.

The theory is still incomplete. Towards improving it, their proponents propose crucial area which needs to be tackled,these includes; (1) formulation of mathematical model for examining the interactionbetween neurons and electromagnetic field, (2) exploration of the interaction between ion channel and brain's electromagnetic field at the quantum level,(3) examination of the role of biological neurons in the fields of information processing, and (4) investigation of the possible role of electromagnetic field inartificial intelligent[117].

## 8. Consciousness as the universal phenomena

According to Upanishidh principle[118] consciousness pervades and illuminates the mind-body, enabling to function as shown in Figure III. It exists even when the mind-body is not there. This approaches also proposes that consciousness is not the part of the mind-body, but it can be known by the function of the mind-body. This approach is contrary to modern neuroscientists, where consciousness is considered to emerge from the functioning of the brain.

Upanishidh approach posits scientific argument which can be explored for further understanding of consciousness. These includes; consciousness as universal phenomena, non-local, non-causal, it is known from the interaction of the brain. It will be interesting to establish a scientific understanding of the consciousness based on the Upanishidh principle. In our future study we will explore this phenomenon.

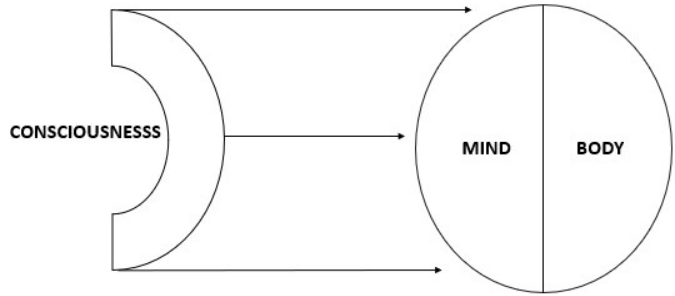

Figure III:Upanishidh illustration of consciousness process

## 9. Outlook

As consciousness has been associated with quantum computation in microtubules, the study of microtubules at the nanoscale is expected to bridge the gap between nanoscience and consciousness. It will bring a new insight of consciousness at the smallest scale. Probably, it will help us to understand why single-cell organisms like paramecium exhibit cognitive behavior although do not possess neurons, synapses, or the brain. Another area of study is that of investigation of existence of operating quantum network architectures in the brain[15]. This approach, will assist in identifying the physically-based quantum mechanism of consciousness. Further, quantum network will lead to quantum memory and communication channel for information processing and transfer. In recent, there is a proposal fora Quantum brain network (QBraiNs), interdisciplinary field integrating knowledge and method from quantum computing, artificial intelligence, and neurobiology[119]. The aim of this is to develop connectivity between the quantum computer and the human brain. It is expected to yield a hybrid classical-quantum networks of wetware and hardware nodes. The quantum field approach has enlightened a better understanding of memory, there is an open question as whether it can also assist towards understanding consciousness.

Although consciousness seems not to speak the language of physical mathematics, but future understanding of consciousness in terms of physical mathematics is of crucial importance. It will enable us to compute the quantity or value of consciousness of a physical system. The achievement of this will help to predict different phenomena of consciousness which are still unclear. For example, what distinguish conscious system from unconscious? Are animals, plant, and artificial intelligent system conscious?Theelectro-chemical transmission of signals does not provide a reasonable explanation of higher brain functions such as consciousness, emotion, learning, and perception. By this we hope probably the study of biophoton as an alternative means of transmission of information in the brain will provide new insight toward understanding consciousness. Furthermore, consciousness also is associated with structure of matter and information. As information requires a channel or network for propagation, the study of the network physics of the brain will enrich the understanding of consciousness. Since the advances of non-invasive imaging techniques allows the comprehensive mapping of structural and functional patterns of the brain. These patterns can be extended to understanding of how the brain support cognitive process.

## Acknowledgements

We thank Indian Council for Cultural Relations for funding. We also extend our grateful thankful for the member of Physics department at Central University of Punjab for their constructive advice.

**Funding:** This study has been supported by Indian Council of Cultural Relations (ICCR) under African Scheme for an author Charles Johnstone to pursue his Ph.D. studies in India.

**Conflict of interest:** The authors declare no competing interests.

**Ethical approval:** No ethical approval is required in this study.

**Informed consent:** Not applicable to this study.

**Author contributions:** PSA: idea, planning, critical revisions CJ: literature search, structuring, drafting